# Reciprocity of thermal diffusion in time-modulated systems

Jiaxin Li[1,2†], Ying Li[3,4,5,1†], Pei-Chao Cao[6†], Xu Zheng[7], Yu-Gui Peng[8], Baowen Li[9,7], Xue-Feng Zhu[6*], Andrea Alù[8*], and Cheng-Wei Qiu[1*]

[1] *Department of Electrical and Computer Engineering, National University of Singapore, Singapore 117583, Singapore*

[2] *School of Mechatronics Engineering, Harbin Institute of Technology, Harbin 150001, China*

[3] *Interdisciplinary Center for Quantum Information, State Key Laboratory of Modern Optical Instrumentation, College of Information Science and Electronic Engineering, Zhejiang University, Hangzhou 310027, China*

[4] *ZJU-Hangzhou Global Science and Technology Innovation Center, Key Lab. of Advanced Micro/Nano Electronic Devices & Smart Systems of Zhejiang, Zhejiang University, Hangzhou 310027, China*

[5] *International Joint Innovation Center, ZJU-UIUC Institute, The Electromagnetics Academy at Zhejiang University, Zhejiang University, Haining 314400, China*

[6] *School of Physics and Innovation Institute, Huazhong University of Science and Technology, Wuhan 430074, China*

[7] *Department of Physics, University of Colorado, Boulder, CO 80309, USA*

[8] *Department of Electrical Engineering, City University of New York, New York, NY 10031, USA*

[9] *Department of Mechanical Engineering, University of Colorado, Boulder, CO 80309, USA*

† These authors contributed equally to this work.
* email: xfzhu@hust.edu.cn ; * email: aalu@gc.cuny.edu ; * e-mail: chengwei.qiu@nus.edu.sg

**The reciprocity principle governs the symmetry in transmission of electromagnetic and acoustic waves, as well as the diffusion of heat between two points in space, with important consequences for thermal management and energy harvesting. There has been significant recent interest in materials with time-modulated properties, which have been shown to efficiently break reciprocity for light, sound, and even charge diffusion. Quite surprisingly, here we show that, from a practical point of view, time modulation cannot generally be used to break reciprocity for thermal diffusion. We establish a theoretical framework to accurately describe the behavior of diffusive processes under time modulation, and prove that thermal reciprocity in dynamic materials is generally preserved by the continuity equation, unless some external bias or special material is considered. We then experimentally demonstrate reciprocal heat transfer in a time-modulated device. Our findings correct previous misconceptions regarding reciprocity breaking for thermal diffusion, revealing the generality of symmetry constraints in heat transfer, and clarifying its differences from other transport processes in what concerns the principles of reciprocity and microscopic reversibility.**

Reciprocity is a fundamental property of wave propagation[1,2] and diffusion[3], implying symmetric field transport in opposite directions. Breaking reciprocity in energy and information transport is essential in components such as diodes, isolators[4-7], rotators[8], rectifiers[9,10], and circulators[11-13], spanning electromagnetics, photonics and acoustic domains. Besides the effects of reciprocal heat transfer in static[14] and moving components[15-19], breaking the symmetry of heat transfer to achieve thermal non-reciprocity[20] is also of great importance for various applications. Devices like heat pipes and thermosyphon diodes are commonly used for thermal management and energy harvesting[21]. In addition, solid-state thermal diodes[22] and rectifiers[23] are the basic elements for thermal information processing in analogy to electronics[24,25].

In general, there are three types of approaches that break reciprocity. The first is to apply an external bias that is an odd function of time under time-reversal symmetry, like magnetic fields or mechanical motion[11,26,27]. For heat transfer, a simple external bias can be realized by introducing mass or energy fluxes that enter and leave the system with a preferred directionality. Such straightforward approach is not very practical, because it usually makes the underlying systems hardly integrable. The second is by using nonlinearity[7,28-30]. Asymmetric thermal conduction has been found in nonlinear materials[31] with temperature-dependent properties such as oxides[32] or shape-memory alloys[33,34], but the reliance on exotic materials limits its applicability and working conditions.

A third approach to break reciprocity, inspired by recent efforts in electromagnetics and acoustics[35,36], has been based on materials with time varying properties. This scheme has received growing interest, since it is easier to be integrated and broadly applicable compared to the first two approaches. The propagation of electromagnetic waves in coupled waveguides has been shown to be non-reciprocal when the electric permittivity $\varepsilon$ is modulated with a travelling wave[35] (Fig. 1a), thanks to asymmetric mode conversions. Similar ideas have been successfully applied to thermal radiation[37] and acoustic waves[38,39]. Interestingly, time modulation can also induce asymmetric transfer of electric charge, which is essentially a diffusive process[40]. Intuitively, this is possible because the governing equation, i.e., Fick's law, contains the same Laplacian term as the wave equation. Different from wave propagation, two material parameters in the diffusion equation— the capacitance $C_e$ and electric conductivity $\sigma$ must be modulated simultaneously (Fig. 1b) to achieve this effect.

Conductive heat transfer in solids is another fundamental diffusive process, whose governing equation (Fourier's law) has the same form as Fick's law. The counterpart of electric conductivity $\sigma$ is thermal conductivity $\kappa$, while the counterpart of capacitance $C_e$ is the product of density and specific heat capacity $\rho c$. In practice, the specific heat capacity $c$ is hardly tunable, so we only consider the modulation of density $\rho$ and thermal conductivity $\kappa$ in the following discussion. It appears quite reasonable to expect thermal non-reciprocity induced by such time modulation[42], considering the continuous success of this approach in electromagnetic[35,36] and acoustic[38,39] wave propagation, and charge diffusio[40]. Here, however, we prove theoretically and present numerical and experimental evidences that it is extremely difficult to break reciprocity in heat transfer using time modulation without resorting to external bias or special materials (Fig. 1c). This result is due to the fact that a time modulation of the density inevitably alters the governing transfer equation by taking into account the necessary mass motion $v$. Our findings indicate that diffusive heat transfer presents inherent constraints that must be carefully treated in its manipulation to break reciprocity.

**Results**

*Diffusion equation under time modulation*

Heat transfer in solids is governed by the diffusion Fourier's law: $\partial(\rho c T)/\partial t = \nabla \cdot (\kappa \nabla T)$, where $T(r,t)$ is the temperature field, $r$ is position vector, and $t$ is time. If the material is linear and not dynamic, the solutions strictly obey reciprocity[41]. In the case of dynamic materials, the density and thermal conductivity vary with time. If both parameters can be freely modulated without introducing additional effects, the Fourier's law becomes

$$\frac{\partial \left[\rho(\bm{r},t)cT\right]}{\partial t} = \nabla \cdot \left[\kappa(\bm{r},t)\nabla T\right] \tag{1}$$

Since Eq. (1) has the same form as the time-modulated Fick's law[40], it is expected that the solution would be in general non-reciprocal[42]. However, as we will discuss in the following, it is impossible to freely modulate the density, since matter that acts as the carrier of thermal energy cannot be created or destroyed. The variation of density $\rho$ must obey the law of mass conservation, which leads to a different governing equation than Eq. (1).

In continuous media, mass conservation is preserved by the continuity equation $\partial\rho/\partial t + \nabla \cdot (\rho\bm{v}) = 0$[43]. If the density varies with time, we inevitably expect mass movement with velocity $\bm{v}$ (Fig. 1c). Since thermal energy is inherent in any material, the movement introduces a convective term $\nabla \cdot (\rho c T \bm{v})$, which does not appear in Eq. (1). This implies that Eq. (1) can hardly be realized within a physical system without providing external energy or mass.

By adding the convective term into Eq. (1), a mass-conserving diffusive heat transfer under time modulation becomes the convection-diffusion process

$$\rho(\bm{r},t)c\frac{\partial T}{\partial t} + \rho(\bm{r},t)c\bm{v}\cdot\nabla T = \nabla \cdot \left[\kappa(\bm{r},t)\nabla T\right], \tag{2}$$

where we assume that there is no other thermal effect and that viscous dissipation is negligible. The detailed effects of the convective term depend on the velocity field $\bm{v}(\bm{r},t)$. In order to focus on the mechanism of time modulation, it is reasonable to only study systems at steady state without externally applied directional mass or energy flux. To be specific, the boundary conditions are constant, and the modulation of material parameters is periodic with time to ensure a stable (time-harmonic) field. No external mass flux exists: $\rho\bm{v}\cdot\bm{n} = 0$, where $\bm{n}$ is the unit normal vector at the system boundary. In time-modulated systems, it is suitable to require instead that the average external mass flux in a time period $t_0$ vanishes

$$\langle \rho \boldsymbol{v} \cdot \boldsymbol{n} \rangle = \int_0^{t_0} \rho(\boldsymbol{r},t)\boldsymbol{v}(\boldsymbol{r},t) \cdot \boldsymbol{n} dt = 0 \tag{3}$$

No accumulated external bias is a central assumption that will be used throughout our analysis.

There are only two types of setups that support density modulation without external bias: in the first case, the density is modulated by mass motion along the heat transfer path (Fig. 1d). The spheres illustrated do not represent microscopic particles but macroscopic components. There can be an exchange of mass between the system and two thermal reservoirs at both ends, but the three parts together restore the original state after a period, hence there is no net directional mass flow. In the second one, the density is modulated by mass entering or leaving the heat transfer path cyclically (Fig. 1e). In the following, we prove that the heat transfer in both setups is inherently reciprocal. We also experimentally build a setup of the second type to validate our prediction.

*Density modulations of the first type*

The first type of modulation scheme follows the one-dimensional (1D) model shown in Fig. 2a. The density $\rho(\chi)$ and thermal conductivity $\kappa(\chi)$ of the material are $d$-periodic functions of $\chi = x - v_0 t$, so their profiles move at constant speed $v_0$ along $x$. According to the 1D continuity equation $\partial \rho / \partial t + \partial (\rho v) / \partial x = 0$, the mass flux $\rho v$ satisfies $\rho v = (\rho - \rho_0)v_0 + C$, where $\rho_0$ is the average density and $C$ is a constant. According to Eq. (3), there is no accumulated mass flux through the system in a time period $t_0 = d/v_0$ (Fig. 2b). Thus, we have $C = 0$ and can solve for the velocity field as $v(\chi) = [\rho(\chi) - \rho_0]v_0/\rho(\chi)$ (See Supplementary Note 1 and Supplementary Fig. 1 for the effects of a nonzero $C$). The 1D heat transfer then obeys

$$\rho(\chi)c\frac{\partial T}{\partial t} + [\rho(\chi) - \rho_0]cv_0\frac{\partial T}{\partial x} = \frac{\partial}{\partial x}\left[\kappa(\chi)\frac{\partial T}{\partial x}\right] \tag{4}$$

The thermal reciprocity (at steady state) is often related to the symmetry between backward and forward heat transfers[20], which are generated by applying fixed temperature boundary conditions $T_0 = T_{cold}$ and $T_L = T_{hot}$ (backward) or $T_0 = T_{hot}$ and $T_L = T_{cold}$ (forward) at the two ends $x = 0$ and $L$, respectively. For Eq. (4), such a symmetry can be proved by comparing the forward and backward heat fluxes. Assuming that $T_f(x,t)$ is the solution for the forward case with the boundary conditions: $T_f(0,t) = T_{hot}$, $T_f(L,t) = T_{cold}$, while $T_b(x,t)$ is the solution for the backward case with the boundary conditions: $T_b(0,t) = T_{cold}$, $T_b(L,t) = T_{hot}$. Given the initial conditions, both solutions should be unique. Their summation $T_t(x,t) = T_f(x,t) + T_b(x,t)$ also satisfies Eq. (4), with initial state $T_t(x,0) = T_f(x,0) + T_b(x,0)$ and boundary conditions $T_t(0,t) = T_t(L,t) = T_{hot} + T_{cold}$. It is easy to check that $T_t(x,t) = T_{hot} + T_{cold}$ is a solution, and must be the unique solution thanks to the uniqueness of $T_f(x,t)$ and $T_b(x,t)$. The heat flux $q(x,t)$ is the sum of conductive and convective heat flux: $q(x,t) = -\kappa(\chi)\partial T/\partial x + [\rho(\chi) - \rho_0]v_0 c T(x,t)$. The heat fluxes $q_f(x,t)$ and $q_b(x,t)$ corresponding to $T_f(x,t)$ and $T_b(x,t)$ then satisfy

$$q_f(x,t) + q_b(x,t) = -\kappa(\chi)\frac{\partial T_t}{\partial x} + [\rho(\chi) - \rho_0]v_0 c T_t(x,t) = [\rho(\chi) - \rho_0]v_0 c(T_{hot} + T_{cold}) \quad (5)$$

Averaging over time gives $\langle q_f(x)\rangle + \langle q_b(x)\rangle = 0$. Considering that the average heat fluxes in and out of the system should balance, we have $\langle q_f(0)\rangle = \langle q_f(L)\rangle = -\langle q_b(0)\rangle = -\langle q_b(L)\rangle$, which meets the condition for a symmetric forward and backward heat transfer: $\langle q_f(0)\rangle = -\langle q_b(L)\rangle$ and $\langle q_f(L)\rangle = -\langle q_b(0)\rangle$, and indicates thermal reciprocity[20].

We can also analytically solve Eq. (4). After a variable change $(x,t)$ to $(\chi = x - v_0 t, \tau = t)$, it is easy to see that Eq. (4) is periodic on $\chi$, so the Bloch theorem applies and gives (see Supplementary Note 1 for details):

$$T(x,t) = e^{-ikx} f(\chi) \quad (6)$$

where $f(\chi)$ is a periodic function with periodicity $d$, and $k$ is the Bloch wavenumber. The temperature solution should be a linear combination of Eq. (6). The dissipative nature of heat transfer indicates that $k$ must have nonzero imaginary part for time-harmonic solutions[44]. Substituting Eq. (6) into (4), the temperature field can be analytically solved using the Fourier series of $f(\chi)$, based on the periodicity of $\rho(\chi)$ and $\kappa(\chi)$.

To verify the solution, we build a 1D model with $d = 1$ cm and $L = 10d$. The density and thermal conductivity are set to be $\rho(\chi) = \rho_0[1 + \Delta_\rho \cos(\beta\chi)]$ and $\kappa(\chi) = \kappa_0[1 + \Delta_\kappa \cos(\beta\chi)]$, where $\rho_0 = 2000$ kg m$^{-3}$, $\Delta_\rho = 0.3$, $\kappa_0 = 100$ W m$^{-1}$ K$^{-1}$, and $\Delta_\kappa = 0.9$. The heat capacity is $c = 1000$ J kg$^{-1}$ K$^{-1}$. We choose two modulation speeds $v_0 = \mu\kappa_0/\rho_0 c$ with $\mu = 1/d$ and $4/d$. Constant temperatures are set as $T_{\text{cold}} = 273$ K and $T_{\text{hot}} = 373$ K to generate a temperature difference $\Delta T = T_{\text{hot}} - T_{\text{cold}} = 100$ K. Our analytical results (solid lines in Fig. 2c and d) are well validated by finite-element simulations (scatter points in Fig. 2c and d) with COMSOL Multiphysics®.

For comparison, we also plot analytical and numerical solutions to the diffusion equation[42]

$$\rho(\chi)c\frac{\partial T}{\partial t} = \frac{\partial}{\partial x}\left[\kappa(\chi)\frac{\partial T}{\partial x}\right] \qquad (7)$$

The solutions to Eq. (4) for the heat transfer are symmetric in the backward (Fig. 2c) and forward (Fig. 2d) directions for all modulation parameters characterized by $\mu$, $\Delta_\kappa$ and $\Delta_\rho$. This is in contrast with the solutions to Eq. (7) with concave/convex profiles (see Supplementary Note 2 for accurate solutions). However, Eq. (7) is a hypothetical one that requires external energy source, because density modulation $\rho(\chi)$ cannot be achieved at no cost. In a virtual system following Eq. (7), there should be a difference between the average heat flux entering and leaving the system at the two ends, showing that additional energy input or extraction is required to compensate it.

### *Density modulations of the second type*

Another way to achieve density modulation without net directional flow is to add/remove matter periodically through the heat transfer path. The simplest setup of such type is the two-dimensional (2D) model shown in Fig. 3a, where the heat transfer path under consideration is the transverse narrow region at $y = 0$. We assume constant temperatures $T_0$ and $T_L$ on the left and right sides at $x = 0$ and $L$, while periodic boundaries are assumed on the upper and lower sides at $y = \pm d_y/2$. $\rho(x,y = 0,t)$ and $\kappa(x,y = 0,t)$ are $d$-periodic functions of $x - v_0 t$, so we consider 2D distributions that are $d$-periodic functions of $\zeta = x + \eta y - v_0 t$ with $\eta = d/d_y$.

Next, we consider mass motion along $y$ with speed $v_y(x,y,t)$ to locally modulate the density. According to the continuity equation $\partial \rho(\zeta)/\partial t + \partial[\rho(\zeta)v_y]/\partial y = 0$, we find $\rho v_y = (\rho - \rho_0)v_{0y} + C$, where $v_{0y} = v_0/\eta$. When $C = 0$, the case is realizable with mass oscillations in $y$ direction, which is almost the same as the 1D case. Here, we apply periodic condition to the upper and lower boundaries and set $C = \rho_0 v_{0y}$, so that $v_y(x,y,t) = v_{0y}$ (Fig. 3a). It is noted that Eq. (3) is still satisfied. The case can be realized on the side surface of a rotating cylinder, which corresponds to the three-dimensional (3D) model discussed later. The 2D heat transfer follows

$$\rho(\zeta)c\frac{\partial T}{\partial t} + \rho(\zeta)cv_{0y}\frac{\partial T}{\partial y} = \frac{\partial}{\partial x}\left[\kappa^x(\zeta)\frac{\partial T}{\partial x}\right] + \frac{\partial}{\partial y}\left[\kappa^y(\zeta)\frac{\partial T}{\partial y}\right] \qquad (8)$$

in which we consider the general case with a thermal conductivity modeled as an anisotropic tensor. Its $xx$ and $yy$ components are $\kappa^x$ and $\kappa^y$, while the off-diagonal components are assumed to be zero. Similar to the 1D case, the solution to Eq. (8) can be solved (see Supplementary Note 3).

A practical 3D setup that realizes such a model is shown in Fig. 3b, which consists of fixed and moving fan-shaped solid plates with density $\rho_A = 8390$ kg m$^{-3}$, heat capacity $c_A = 375$ J kg$^{-1}$ K$^{-1}$, and thermal conductivity $\kappa_A = 123$ W m$^{-1}$ K$^{-1}$. Each plate spans $\pi/2$ with inner and exterior radius $R_1 = 1$ cm and $R_2 = 2$ cm, and thickness $\delta = 0.25$ cm $= d/16$. The total length of the system $L = 5d = 20$ cm. Temperature boundary conditions $T_0 = T_{\text{cold}}$ and $T_L = T_{\text{hot}}$ (backward) or $T_0 = T_{\text{hot}}$ and $T_L = T_{\text{cold}}$ (forward) are applied at $x = 0$ and $L$, respectively, with constant temperatures set as $T_{cold}$ = 273 K and $T_{hot}$ = 323 K. All other boundaries are thermally insulated. Naturally, we can regard the heat transfer path as the portion of the system contained in the region ($[R_1,R_2]$, $[\pi/4,3\pi/4]$, $[0,L]$) of the cylindrical coordinate system ($r, \theta, x$), through which most of the heat flux is conducted. Along the $x$-direction, each moving plate is $\pi/4$ ahead of the previous moving one. All of them rotate at angular speed $-\Omega = -0.03 \times 2\pi$ rad s$^{-1}$. As the moving plates enter or leave the heat transfer path at $y = 0$, the density $\rho(x,y = 0,t)$ and thermal conductivity $\kappa(x,y = 0,t)$ are effectively modulated (see Supplementary Note 3 for the mapping of the 3D model to 2D).

We perform numerical simulations on the 2D and 3D models in Fig. 3a and b. As expected, reciprocal heat transfer is confirmed for all considered temperature distributions along the line at ($r = R_2$, $\theta = \pi/2$) (Fig. 3c,d). The 2D (lines) and 3D (scatter points) numerical results are in good agreement. The temperature distributions on the plate surfaces are also plotted in the insets (see Supplementary Movie 1 for the evolution in time). Reciprocity can be easily shown from the symmetric temperature profiles and the analysis for forward and backward heat fluxes similar as in the first modulation type. The reason for the absence of non-reciprocity resides in the fact that in these diffusive systems the density variations are averaged out over time due to the conservation of mass, and hence there is effectively no density modulation unless there is directional mass transfer. To further illustrate this point, we performed simulations on a hypothetical model where the density of the moving plates is artificially modulated as $\rho_A(t) = \rho_A\{1 + \cos[2\Omega t − (n − 1)\pi/4]\}/2$, for the $n$-th layer of moving plates. The temperature distributions in Fig. 3e and f show concave and convex profiles, indicating non-reciprocity (see Supplementary Movie 2 for the evolution).

*Experimental verification of reciprocal heat transfer in a time-varying material*

We have implemented the 3D geometry in Fig. 3b, as shown in Fig. 4a. The system is built with fan-shape plates attached to a fixed beam and a shaft revolved by a low-speed motor to generate a right-handed rotating profile. The plates are made of brass with the same material properties and shape as in the 3D simulations. The supporting beam and shaft are made of nylon with thermal conductivity 0.3 W m$^{-1}$ K$^{-1}$. The heat source and sink are water baths at 323 K and 273 K which are made in contact with both ends of the system via two copper plates and generate a temperature difference $\Delta T = T_{\text{hot}} - T_{\text{cold}} = 20$ K. Different from the numerical simulations, a 0.2 mm gap is made between each adjacent plates, which is filled with thermal grease (thermal conductivity 4.38 W m$^{-1}$ K$^{-1}$) for conduction and silicone oil (thermal conductivity 0.16 W m$^{-1}$ K$^{-1}$) for lubrication.

Therefore, the interface thermal resistance is nonnegligible. In addition, the natural convective heat exchange with the ambient air introduces another term $h(T_\infty - T)$, where $h$ is the heat exchange rate and $T_\infty$ is the room temperature. Due to these two factors, the temperature gradients at the center are much smaller compared to the linear profile (Fig. 4b,c). However, the two factors do not have any directionality, so the temperature profiles for the backward (Fig. 4b) and forward (Fig. 4c) cases are still symmetric, demonstrating reciprocal heat transfer in the time-modulated system. See Supplementary Movie 3 for videos of the rotating device and the temperature profiles. Reciprocity is further confirmed by the temperature distribution in the top portion of the system (dashed green line in Fig. 4a), where the fixed plates are placed (Fig. 4d). In Fig. 4d, to compare with the experimental results, numerical simulations were performed on the 3D model with interface thermal resistance (0.2 mm thickness and 1 W m$^{-1}$ K$^{-1}$ thermal conductivity) and natural air convection ($h = 10$ W m$^{-2}$ K$^{-1}$, $T_\infty = T_{\text{cold}} + 0.485\Delta T$). $T_\infty$ was chosen to meet the plateau in the experimental data. The measured curves for backward and forward heat transfer are not only symmetric, but also in good agreement with the simulated results.

**Discussion**

In this paper, we have shown that time-modulated density and thermal conductivity generally does not break thermal reciprocity in practice. For other processes governed by the momentum equation (wave propagation) or the continuity equation (charge diffusion), the driving approach of time modulation can be decoupled from the field. For heat transfer, however, time modulation not only *indirectly* influences the thermal field through the density, but also inevitably and *directly* influences the field through the convective term. Such effects counter-balance each other, and reciprocity is preserved.

While we did not explicitly discuss the case of a varying specific heat capacity $c$, it is easy to recognize that our arguments equally apply to this scenario. First of all, if the variation is simply a consequence of mechanical motions, the material derivative of $\rho c$ must be zero, which gives $\partial(\rho c)/\partial t + \nabla \cdot (\rho c \boldsymbol{v}) = 0$. Therefore, Eq. (2) and the following analysis still apply. Second, if the specific heat capacity of the material can really be modulated at will, it is possible to avoid the convective term and thereby in principle achieve non-reciprocity. However, this is only possible in very limited scenarios, such as using caloric materials[45]. Even if one can break thermal reciprocity by doing so, it is arguably less convenient than using nonlinearities, without the need for time modulation.

Our work suggests that thermal reciprocity has more fundamental resilience than other transport mechanisms. These findings may have important implications for the design of thermal devices and other dissipative wave propagation systems.

**Acknowledgements**

C.-W.Q. acknowledges the support from Ministry of Education, Singapore (Grant No.: R-263-000-E19-114). X.-F.Z. and P.-C.C. acknowledge the financial support of National Natural Science

Foundation of China (Grant Nos. 11674119, 11690030, and 11690032), and the Bird Nest Plan of HUST. J.L. acknowledges the China Scholarship Council (CSC) for financial support. A.A. acknowledges the support of the Air Force Office of Scientific Research and the Department of Defense.

**Author contributions**

J.L., Y.L., and C.-W.Q. conceived the idea. J.L. proposed the analytical model. J.L. and Y.L. performed the theoretical derivations. J.L. and Y.L. designed and performed the numerical simulations. J.L., Y.L., and P.-C.C. designed and performed the experiments. X.Z., B.L. and A.A. contributed to the analysis on heat flux. Y.L. wrote the manuscript. X.-F.Z., A.A. and C.-W.Q. supervised the work. All the authors discussed and contributed to the manuscript writing.

**Competing financial interests**

The authors declare no competing financial interests.

**Data availability**

The data that support the findings of this study are available from the corresponding author (C.-W.Q.) on request.

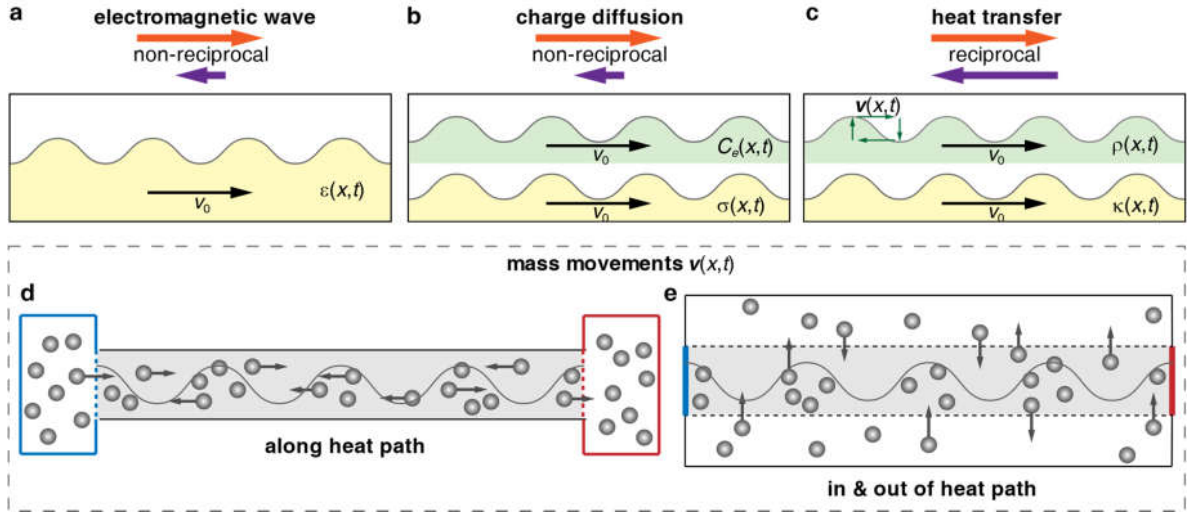

**Figure 1. Transport processes in dynamic materials under time modulation. a** Non-reciprocal propagation of electromagnetic wave can be induced by spatial-temporally modulating the electric permittivity $\varepsilon(x,t)$ as a travelling wave (with speed $v_0$). **b** Non-reciprocal diffusion of electric charges can be induced by spatial-temporally modulating the capacitance $C_e(x,t)$ and electric conductivity $\sigma(x,t)$. **c** The reciprocity of heat transfer cannot be broken by modulating the density $\rho(x,t)$ and thermal conductivity $\kappa(x,t)$ since it is preserved by the continuity equation. Following the law, mass movements $v(x,t)$ (green arrows) must exist to achieve density modulation. **d-e** Two types of mass movements (dark grey spheres): **d** along the heat transfer path (grey region). **e** moving in and out of the heat transfer path.

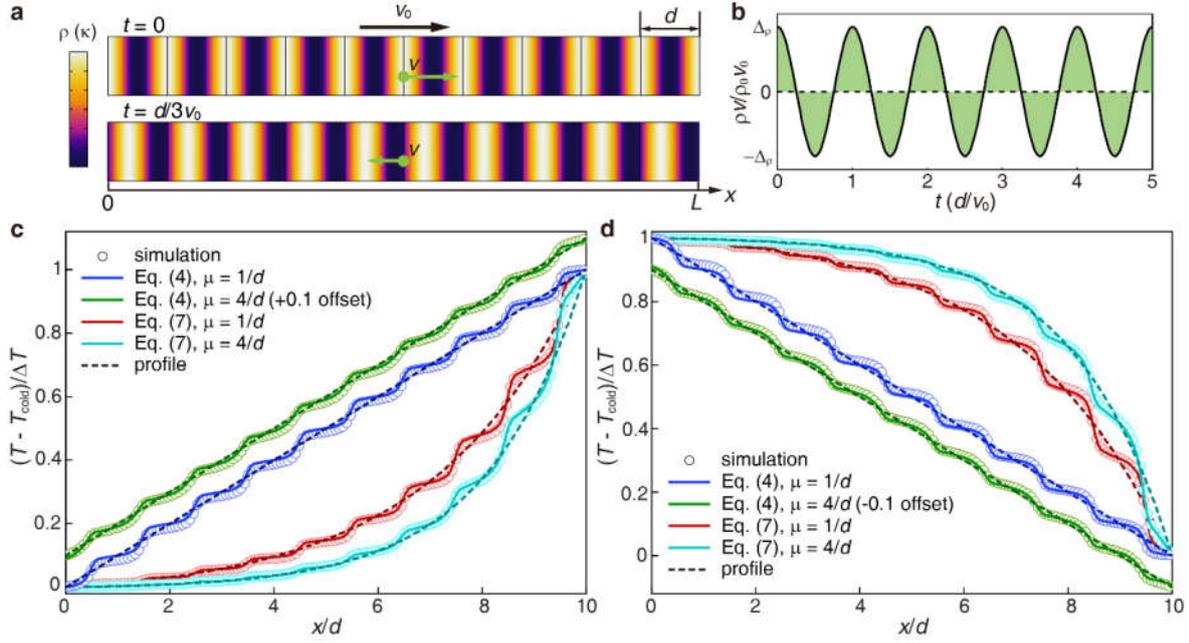

**Figure 2. Heat transfer under 1D density modulation. a** Density $\rho$ and thermal conductivity $\kappa$ (color maps) move as travelling waves at speed $v_0$ with wavelength $d$. To achieve density modulation, actual mass movements (green arrows) at speed $v$ must exist. **b** The total mass flux in a time period should be zero to keep a cyclic and close setup. **c-d** Backward (**c**) and forward (**d**) temperature distributions of the system (Eq. (4)) at $t = 300d/v_0$, compared with those of a virtual system without mass movements (Eq. (7)). Scatter points are simulated results, lines are analytical results, dashed lines are analytical solutions of the homogenized profiles. For clarity, the results of Eq. (4) at modulating speed $\mu = 4/d$ are shifted to have ±0.1 offsets.

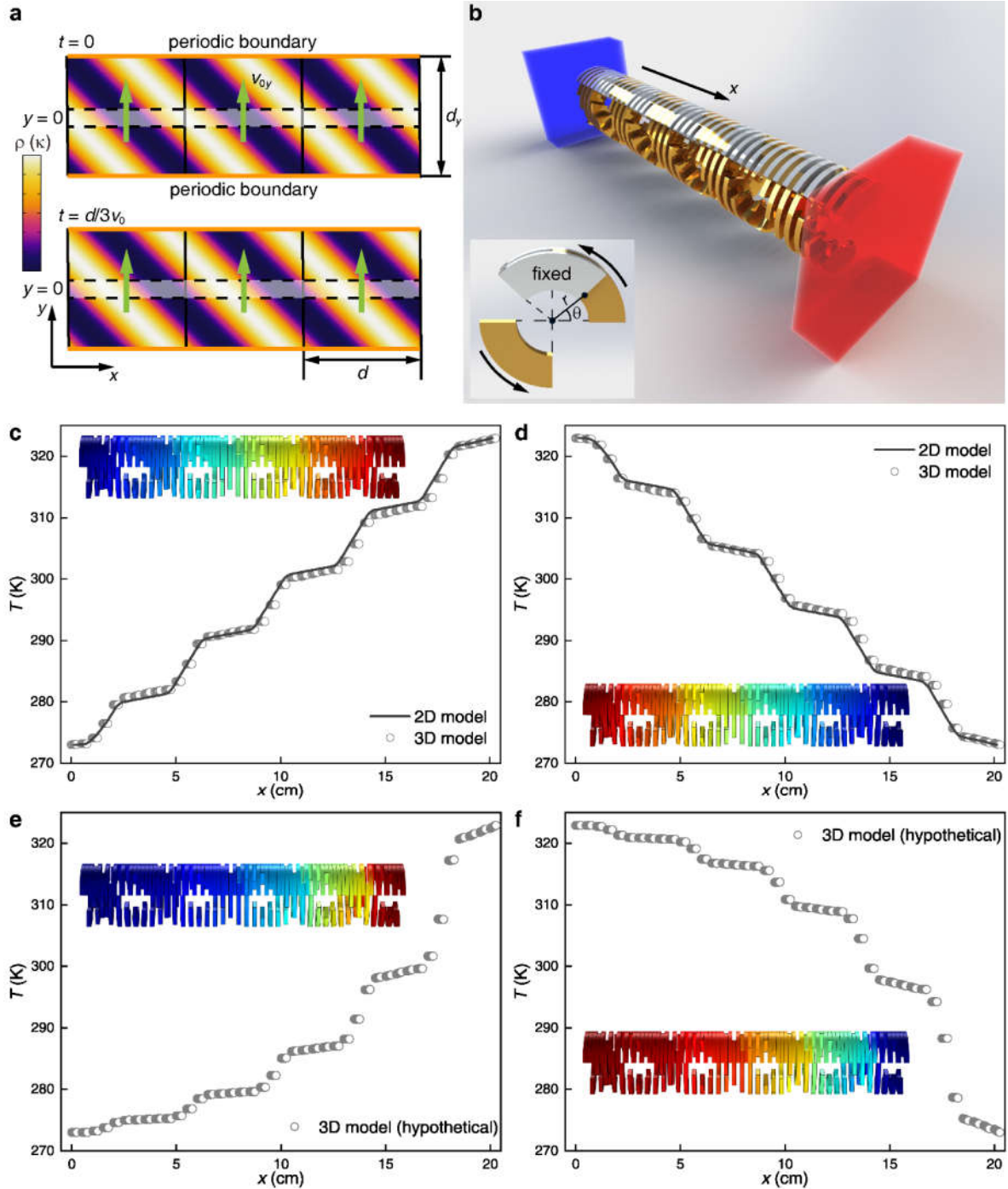

**Figure 3. Heat transfer under 2D and 3D density modulation. a** Density $\rho$ and thermal conductivity $\kappa$ profiles (color maps) of 2D travelling waves such that the profiles move at speed $v_0$ in $x$-direction (marked by dashed lines). The density is modulated by mass movement in $y$-

direction at speed $v_{0y}$ (green arrows). The upper and lower boundaries are periodic. **b** A three-dimensional (3D) model in a $(r,\theta,x)$ cylindrical coordinate system. Similar $\rho$ and $\kappa$ modulations are achieved with fixed (white) and moving fan-shaped plates (yellow). **c-f** Backward (**c,e**) and forward (**d,f**) temperature distributions on the top line $r = R_2$, $\theta = \pi/2$ extracted from simulated results for the 2D (lines) and 3D (scatters) models. The insets show the entire temperature distributions. The results in **e** and **f** use a hypothetical 3D model where the moving plates have time-varying masses, violating the law of mass conservation.

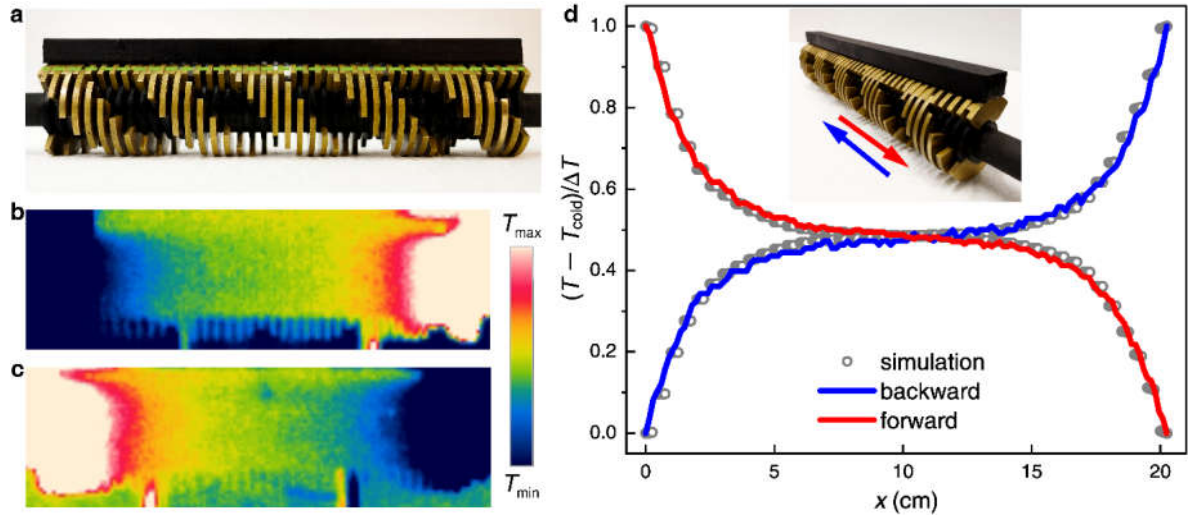

**Figure 4. Experimental verification of reciprocal heat transfer under time-modulation. a** Photo of the system built according to the 3D modulation model in Fig. 3. The movable brass plates are rotated to generate a right-ward moving profile. **b-c** Temperature profiles for (**b**) backward and (**c**) forward heat transfer. The color range is fixed to clearly show the small temperature variation. **d** Experimentally measured temperature profiles along the top of the system (dashed green line in **a**). The scatter points refer to simulated results considering natural air convection and interface thermal resistance. Inset shows the photo of the system from an oblique view.

**Supplementary Note 1: Analytical model and solution of Eq. (4)**

*1D analytical model*

A mass-conserving diffusive 1D heat transfer system under time modulation is governed by the convection-diffusion equation

$$\rho(x,t)c\frac{\partial T}{\partial t}+\rho(x,t)c v\cdot\nabla T=\nabla\cdot\left[\kappa(x,t)\nabla T\right] \quad (S9)$$

where the density $\rho$ and thermal conductivity $\kappa$ of the material are $d$-periodic functions of $\chi = x - v_0 t$. Based on the periodicity of $\rho(\chi)$ and $\kappa(\chi)$, we write their expanded Fourier series as

$$\rho(\chi)=\sum_l \rho_l e^{-il\beta\chi} \quad (S10)$$

$$\kappa(\chi)=\sum_l \kappa_l e^{-il\beta\chi} \quad (S11)$$

where the integer index $l$ takes $0, \pm1, \pm2, \ldots$ and $\beta = 2\pi/d$. The coefficients satisfy $\rho_l = \rho_{-l}^*$ and $\kappa_l = \kappa_{-l}^*$ for the reality of the parameters, where the superscript star means complex conjugation. By substituting $\rho(\chi)$ into the continuity equation

$$\frac{\partial\rho}{\partial t}+\frac{\partial(\rho v)}{\partial x}=0 \quad (S12)$$

we obtain

$$\frac{\partial(\rho v)}{\partial x}=\sum_l(-il\beta)v_0\rho_l e^{-il\beta\chi} \quad (S13)$$

$$\rho v = v_0\sum_l \rho_l e^{-il\beta\chi}+g(t)=\rho v_0+g(t) \quad (S14)$$

where $g(t)$ is $d/v_0$-periodic in order to reach stable state. Without loss of generality, we neglect this additional time dependence and simply let $g(t) = -\rho_0 v_0 + C$, which does not affect any of the conclusions. Then

$$\rho v = (\rho - \rho_0)v_0 + C \quad (S15)$$

If we require there is no accumulated net mass flux through the system, the average of $\rho v$ over a period of time should be 0, which gives $C = 0$. Otherwise, $C \neq 0$ means there exists a directional mass flux through the system.

Substituting Eq. (S15) into Eq. (S9) gives the general 1D convection-diffusion equation

$$\rho(\chi)c\frac{\partial T}{\partial t}+\left\{\left[\rho(\chi)-\rho_0\right]v_0+C\right\}c\frac{\partial T}{\partial x}=\frac{\partial}{\partial x}\left[\kappa(\chi)\frac{\partial T}{\partial x}\right] \quad (S16)$$

When $C = 0$, it is the case we discussed in the main text. When $C \neq 0$, there is additional constant mass flux $C$ passing through the system, which could induce non-reciprocity.

*Analytical solution of the general 1D convection-diffusion equation*

After a variable change $(x,t)$ to $(\chi = x - v_0 t, \tau = t)$, it is easy to see that Eq. (S16) is periodic on $\chi$, so the Bloch theorem applies and gives a solution

$$T(x,t) = e^{-ik\chi} e^{i\Omega \tau} f(\chi) = e^{-ikx} e^{i(\Omega + kv_0)t} f(\chi) \tag{S17}$$

where $\Omega$ is the frequency and $k$ is the Bloch wavenumber. Eq. (S17) describes a temperature profile characterized by $\Omega$ and $k$, and modulated by a $d$-periodic function $f(\chi)$. For time-harmonic solutions, the temperature profile should not vary with time, so $\Omega + kv_0 = 0$. Then the solution at harmonic steady state becoms

$$T(x,t) = e^{-ikx} f(\chi) \tag{S18}$$

Then the temperature solution should be a linear combination of Eq. (S18). For the convenience of following calculations, we rewrite the Bloch wavenumber as $k = \alpha i$ and just consider the case that $\alpha$ is a real number. When $\alpha \neq 0$, we obtain

$$T(x,t) = T_0 + \Delta T \frac{e^{\alpha x} f(\chi) - f(0)}{e^{\alpha L} f(L) - f(0)} = C_1 e^{\alpha x} f(\chi) + C_2 \tag{S19}$$

When $\alpha = 0$, the solution to Eq. (S19) is undetermined. To resolve the problem, we specify the solution is in a form with a linear profile when the analytical solution of $\alpha$ is $\alpha = 0$. That is

$$T(x,t) = C_1 [x + f(\chi)] + C_2 \tag{S20}$$

In Eqs.(S19) and (S20), $C_1$ and $C_2$ are constants, $f(\chi)$ is a $d$-periodic function that can be expanded as

$$f(\chi) = \sum_m F_m e^{-im\beta\chi} \tag{S21}$$

where the integer index $m$ takes $0, \pm 1, \pm 2, \ldots$ and $F_0 = 1$. The coefficients satisfy $F_m = F_{-m}^*$ for the reality of the temperature. The boundary conditions are not satisfied by the oscillatory Eq. (S19) or (S20), but a good approximation can be made by requiring the boundary conditions to be met at $t = 2\pi n d/v_0$, $n = 0, 1, 2\ldots$

By substituting Eq. (S19) into Eq. (S16), we have

$$\kappa f'' + (2\alpha\kappa + \kappa' + \rho_0 c v_0 - C) f' + [\alpha\kappa + \kappa' - (\rho - \rho_0) c v_0 - C] \alpha f = 0 \tag{S22}$$

We can Fourier expand the simplified equation using Eqs.(S10), (S11) and (S21), and write the $n$th order of Eq. (S22) as

$$\sum_m \left[ (\alpha - i\beta n)(\alpha - i\beta m)\kappa_{n-m} - \alpha \rho_{n-m} c v_0 + \delta_{n,m}(\alpha - i\beta n)(\rho_0 v_0 - C)c \right] F_m = 0 \quad \text{(S23)}$$

where $n = 0, \pm 1, \pm 2, \ldots$ and $\delta_{nm}$ is the Kronecker delta. Then the solutions to Eq. (S22) is equivalent to that of the matrix equation as following

$$[G_{n,m}][F_m] = [0] \quad \text{(S24)}$$

where $[G_{n,m}]$ is defined as

$$G_{n,m} = (\alpha - i\beta n)(\alpha - i\beta m)\kappa_{n-m} - \alpha \rho_{n-m} c v_0 + \delta_{n,m}(\alpha - i\beta n)(\rho_0 v_0 - C)c \quad \text{(S25)}$$

The equation can be solved with enough accuracy by cutting to the 4th order (setting $F_m = 0$ for $|m| > 4$). Then $\alpha$ and $F_m$ ($m = \pm 1, \pm 2, \pm 3, \pm 4$) can be can be solved numerically. Combining the temperature boundary conditions at both ends to determine the two coefficients $C_1$ and $C_2$ in Eq. (S19), we can easily obtain the approximation of temperature distribution. During the solving process, it can be found that when $C \neq 0$, $\alpha$ has a nonzero real solution, which indicates a concave/convex temperature profile. However, when $C = 0$ (the case discussed in the main text), the real solution of $\alpha$ is $\alpha = 0$, indicating a liner temperature profile in the form of Eq. (S20). Therefore, the heat transfer in materials with the first type of density modulation (Fig. 1d of the main text) and without net directional flow of mass is reciprocal. In this case ($C = 0$), its analytical solution can be solved by substituting Eq. (S20) into Eq. (S16) and performing similar process as above.

We adopt the distributions of $\rho$ and $\kappa$ as

$$\rho(\chi) = \rho_0 \left[ 1 + \Delta_\rho \cos(\beta\chi) \right] \quad \text{(S26)}$$

$$\kappa(\chi) = \kappa_0 \left[ 1 + \Delta_\kappa \cos(\beta\chi) \right] \quad \text{(S27)}$$

and perform COMSOL simulations to verify the accuracy of the analytic solution. The parameters are set as $\rho_0 = 2000$ kg m$^{-3}$, $c = 1000$ J kg$^{-1}$ K$^{-1}$, $\kappa_0 = 100$ W m$^{-1}$ K$^{-1}$ and $d = 1$ cm. There are two typical values of $C$, namely 0 and $\rho_0 v_0$ that correspond to cyclic mass movement and uniform motion at constant speed $v_0$, respectively. Therefore, we calculate solutions for $C/\rho_0 v_0 = 0, 0.1, 0.2, 0.5$, and 1 to cover different levels of average mass flux through the system (Supplementary Fig. 1a). The backward (Supplementary Fig. 1b) and forward (c) temperature distributions are obtained with $\Delta_\rho = 0.3$, $\Delta_\kappa = 0.9$, and $\mu = \rho_0 v_0/\kappa_0 = 1/d$. All the lines in Supplementary Fig. 1 are analytical solutions, while the scatters are the results of COMSOL simulations. We also see that the analytical solutions are very accurate even for extremely curved profiles and confirm the non-reciprocal heat transfer for $C \neq 0$.. To illustrate the nature of the non-reciprocity, we plot the dependence of $\alpha$ on the modulating speed $\mu$, amplitudes of modulation $\Delta_\kappa$ and $\Delta_\rho$ in Fig. S1d, e, and f, respectively. For $C \neq 0$, it is easy to see that the non-reciprocity is not generated by time modulation, because $\alpha$ is nonzero at $\mu = 0$, $\Delta_\kappa = 0$, and $\Delta_\rho = 0$. On the contrary, $\alpha$ is actually maximized at $\mu = 0$ and $\Delta_\rho = 0$ (Supplementary Fig. 1d and f).

**Supplementary Note 2: Analytical solution of Eq. (7)**

We also use the temperature distribution in the form as Eq. (S19) to solve the 1D hypothetical diffusion equation

$$\rho(\chi) c \frac{\partial T}{\partial t} = \frac{\partial}{\partial x}\left[\kappa(\chi)\frac{\partial T}{\partial x}\right] \tag{S28}$$

Substituting Eq. (S19) into (S28) gives

$$\kappa f'' + (2\alpha\kappa + \kappa' + \rho c v_0) f' + (\alpha\kappa + \kappa')\alpha f = 0 \tag{S29}$$

Using the Fourier expansions of the periodic functions $\rho$, $\kappa$ and $f$ in Eqs. (S10), (S11) and (S21), we can write the $n$th order of Eq. (S29) as

$$\sum_m \left[(\alpha - i\beta n)(\alpha - i\beta m)\kappa_{n-m} - i\beta m \rho_{n-m} c v_0\right] F_m = 0 \tag{S30}$$

where $n = 0, \pm 1, \pm 2, \ldots$ Then the solutions to Eq. (S29) is equivalent to that of the matrix equation as following

$$[H_{n,m}][F_m] = [0] \tag{S31}$$

where $[G_{n,m}]$ is defined as

$$G_{n,m} = (\alpha - i\beta n)(\alpha - i\beta m)\kappa_{n-m} - i\beta m \rho_{n-m} c v_0 \tag{S32}$$

We follow the same procedure as for Eq. (S24) to solve Eq. (S31) up to the fourth order. By numerically solving $\alpha$ and $F_m$ ($m = \pm 1, \pm 2, \pm 3, \pm 4$) from Eq. (S31) we can obtain the temperature solution. The results are in good agreement with numerical simulations as shown in the main text. Under the settings there, we found that maximum $\alpha$ is reached at $\mu = 3.2/d$.

For the conductive heat flux is the only constituent of energy flux in the system, we have the total heat flux $q(x,t)$ as following based on the analytical temperature solution

$$q(x,t) = -C_1 \sum_{n,m} \kappa_{n-m} (\alpha - i\beta m) F_m e^{\alpha x - in\beta \chi} \tag{S33}$$

Clearly, the average of heat flux over time is the the zeroth order of Eq. (S33), that is

$$\langle q(x) \rangle = -C_1 \sum_{m} \kappa_{-m} (\alpha - i\beta m) F_m e^{\alpha x} \tag{S34}$$

It is noted that at the two ends of the system, the average heat fluxes $\langle q(0) \rangle$ and $\langle q(L) \rangle$ are obviously not equal. The difference clearly showes that additional energy input or extraction is required to compensate it, which is also hard to implement in practice.

**Supplementary Note 3: Analytical model and solution of Eq. (8)**

*Analytical solution of the general 2D convection-diffusion equation*

For the second type of modulation, we consider mass motion along $y$ with speed $v_y(x,y,t)$ to modulate locally the density. The 2D convection-diffusion satisfies

$$\rho(\zeta) c \frac{\partial T}{\partial t} + \rho(\zeta) c v_y \frac{\partial T}{\partial y} = \frac{\partial}{\partial x}\left[\kappa^x(\zeta) \frac{\partial T}{\partial x}\right] + \frac{\partial}{\partial y}\left[\kappa^y(\zeta) \frac{\partial T}{\partial y}\right] \tag{S35}$$

where $\zeta = x + \eta y - v_0 t$. According to the continuity equation $\partial \rho(\xi)/\partial t + \partial [\rho(\xi) v_y]/\partial y = 0$, we find $\rho v_y = (\rho - \rho_0) v_{0y} + C$, where $v_{0y} = v_0/\eta$. Substituing it into Eq. (S35), we have the general 2D convection-diffusion equation

$$\rho(\zeta) c \frac{\partial T}{\partial t} + \{[\rho(\zeta) - \rho_0] v_{0y} + C\} c \frac{\partial T}{\partial y} = \frac{\partial}{\partial x}\left[\kappa^x(\zeta) \frac{\partial T}{\partial x}\right] + \frac{\partial}{\partial y}\left[\kappa^y(\zeta) \frac{\partial T}{\partial y}\right] \tag{S36}$$

Consider the general form of the solution

$$T(x,y,t) = C_1 e^{\alpha x} f(\zeta) + C_2 \tag{S37}$$

Substuting Eq. (S37) into (S36) gives

$$\left(\kappa^x + \eta^2 \kappa^y\right) f'' + \left[2\alpha \kappa^x + \kappa^{x\prime} + \eta^2 \kappa^{y\prime} + (\rho_0 v_0 - \eta C) c\right] f' + \left(\alpha \kappa^x + \kappa^{x\prime}\right) \alpha f = 0 \tag{S38}$$

The $n$th order of its Fourier expansion is

$$\sum_m \left[(\alpha - i\beta n)(\alpha - i\beta m) \kappa^x_{n-m} - \beta n m \eta^2 \kappa^y_{n-m} - \delta_{n,m} i\beta n (\rho_0 v_0 - \eta C) c\right] F_m = 0 \tag{S39}$$

If $\rho$ and $\kappa$ are specified, Eq. (S39) can be solved to higher orders to get more detailed results of $\alpha$ and $F_m$ with the same technique as above. During the calculation, it can be found that $\alpha = 0$ is the only real solution and independent of $C$, which shows that the general 2D convection-diffusion system is reciprocal. Especially when $C = \rho_0 v_{0y}$ (the case in the main text), with the property of Eq. (S19) and (S20): $v_{0y}\partial T/\partial y = -\partial T/\partial t$, Eq. (S36) can be transformed to

$$\frac{\partial}{\partial x}\left[\kappa^x(\zeta)\frac{\partial T}{\partial x}\right] + \frac{\partial}{\partial y}\left[\kappa^y(\zeta)\frac{\partial T}{\partial y}\right] = 0 \tag{S40}$$

Since the density disappears in Eq. (S40), the modulation of thermal conductivity alone is insufficient to generate thermal non-reciprocity. This analysis clearly demonstrates how this class of density modulation also cannot break reciprocity.

### *2D model mapping the 3D implementation*

For convenience, we assume that the 3D setup (density $\rho_A = 8390$ kg m$^{-3}$, heat capacity $c_A = 375$ J kg$^{-1}$ K$^{-1}$, and thermal conductivity $\kappa_A = 123$ W m$^{-1}$ K$^{-1}$) is put in air (density $\rho_B = 1.3$ kg m$^{-3}$, heat capacity $c_B = 1016$ J kg$^{-1}$ K$^{-1}$, and thermal conductivity $\kappa_B = 0.025$ W m$^{-1}$ K$^{-1}$). By projecting the side surface $r = R_2$ of the 3D model onto a plane $(x, y)$, we obtain the 2D thermal conductivity distribution on a slice $2n\delta \leq x \leq 2(n + 1)\delta$, $-\pi R_2/4 \leq y \leq \pi R_2/4$ containing a fixed plate and its adjacent region, where $n = 0, 1, \ldots$ The density times heat capacity and thermal conductivity of the fixed plate are always

$$\rho c_{\text{fix}}(x,y,t) = \rho_A c_A \tag{S41}$$

$$\kappa_{\text{fix}}(x,y,t) = \kappa_A \tag{S42}$$

For the adjacent region where the moving plates could enter and leave, the density times heat capacity and thermal conductivity are square wave distributions

$$\rho c_{\text{mov}}(x,y,t) = \rho_B c_B + (\rho_A c_A - \rho_B c_B)\text{rect}(\zeta) \tag{S43}$$

$$\kappa_{\text{mov}}(x,y,t) = \kappa_B + (\kappa_A - \kappa_B)\text{rect}(\zeta) \tag{S44}$$

where $\zeta = x + \eta y - v_0 t$, $\eta = 16\delta/(\pi R_2)$, and $v_0 = \Omega R_2 \eta$. rect($\zeta$) is a square wave with wavelength $d = 16\delta$. To simplify the distributions, we propose that the effective density times heat capacity and the anisotropic effective thermal conductivity of two adjacent slices can be estimated as

$$\rho c(\zeta) = (\rho c_{\text{fix}} + \rho c_{\text{mov}})/2 \tag{S45}$$

$$\kappa^x(\zeta) = 2\left(\kappa_{\text{fix}}^{-1} + \kappa_{\text{mov}}^{-1}\right)^{-1} \tag{S46}$$

$$\kappa^y(\zeta) = (\kappa_{\text{fix}} + \kappa_{\text{mov}})/2 \tag{S47}$$

Note that many orders of the Fourier series are required to accurately describe a square wave, so we directly performed numerical simulations of the 2D model with the distributions in Eq. (S45)-(S47), instead of calculating the analytical solutions. The temperature field inside the heat transfer path meets well with that of the 3D model. As confirmed by the numerical results in Fig. 3c and d (lines) of the main text.

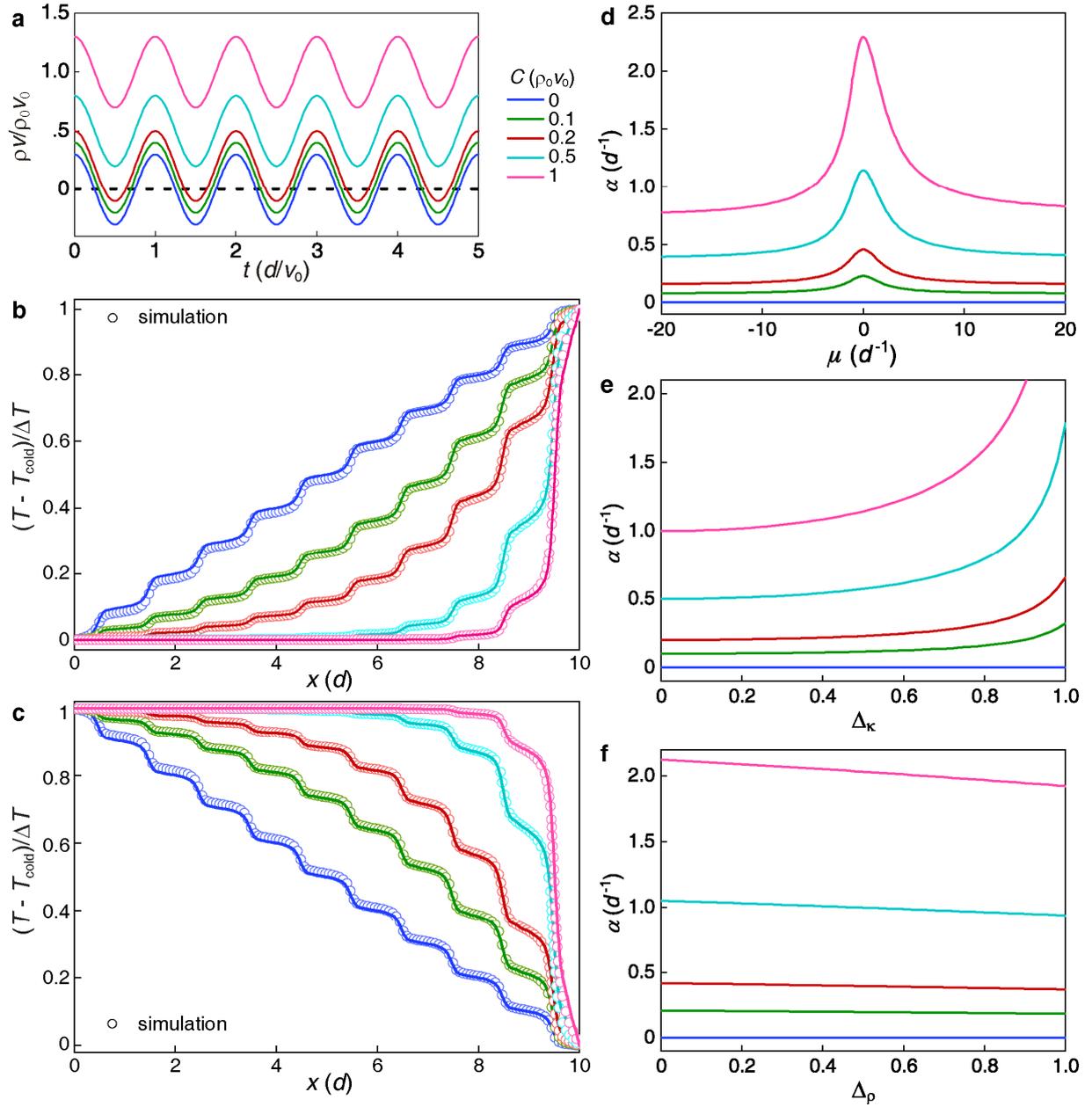

**Supplementary Figure 1. One-dimensional heat transfer under density modulation of the first type and with nonzero average mass flux. a** For $C \neq 0$, there is an additional mass flux through the system. **b-c** Backward (**b**) and forward (**c**) temperature distributions. The scatters are results of numerical simulations. **d-f** The asymmetry of temperature distributions (characterized by $\alpha$) for different modulation speed $\mu$ (**d**), and amplitudes of modulation $\Delta_\kappa$ (**e**) and density $\Delta_\rho$ (**f**).